\documentclass[a4paper]{jpconf}
\usepackage{graphicx}
\usepackage{subfigure}
\usepackage{amsmath}
\usepackage{amsfonts}
\usepackage{amssymb}
\begin{document}
\let\normalcolor\relax


\newcommand{\drawsquare}[2]{\hbox{%
\rule{#2pt}{#1pt}\hskip-#2pt
\rule{#1pt}{#2pt}\hskip-#1pt
\rule[#1pt]{#1pt}{#2pt}}\rule[#1pt]{#2pt}{#2pt}\hskip-#2pt
\rule{#2pt}{#1pt}}

\newcommand{\Yfund}{\raisebox{-.5pt}{\drawsquare{6.5}{0.4}}}
\newcommand{\Ysymm}{\raisebox{-.5pt}{\drawsquare{6.5}{0.4}}\hskip-0.4pt%
        \raisebox{-.5pt}{\drawsquare{6.5}{0.4}}}
\newcommand{\Ythrees}{\raisebox{-.5pt}{\drawsquare{6.5}{0.4}}\hskip-0.4pt%
          \raisebox{-.5pt}{\drawsquare{6.5}{0.4}}\hskip-0.4pt%
          \raisebox{-.5pt}{\drawsquare{6.5}{0.4}}}
\newcommand{\Yfours}{\raisebox{-.5pt}{\drawsquare{6.5}{0.4}}\hskip-0.4pt%
          \raisebox{-.5pt}{\drawsquare{6.5}{0.4}}\hskip-0.4pt%
          \raisebox{-.5pt}{\drawsquare{6.5}{0.4}}\hskip-0.4pt%
          \raisebox{-.5pt}{\drawsquare{6.5}{0.4}}}
\newcommand{\Yasymm}{\raisebox{-3.5pt}{\drawsquare{6.5}{0.4}}\hskip-6.9pt%
        \raisebox{3pt}{\drawsquare{6.5}{0.4}}}
\newcommand{\Ythreea}{\raisebox{-3.5pt}{\drawsquare{6.5}{0.4}}\hskip-6.9pt%
        \raisebox{3pt}{\drawsquare{6.5}{0.4}}\hskip-6.9pt
        \raisebox{9.5pt}{\drawsquare{6.5}{0.4}}}
\newcommand{\Yfoura}{\raisebox{-3.5pt}{\drawsquare{6.5}{0.4}}\hskip-6.9pt%
        \raisebox{3pt}{\drawsquare{6.5}{0.4}}\hskip-6.9pt
        \raisebox{9.5pt}{\drawsquare{6.5}{0.4}}\hskip-6.9pt
        \raisebox{16pt}{\drawsquare{6.5}{0.4}}}
\newcommand{\Yadjoint}{\raisebox{-3.5pt}{\drawsquare{6.5}{0.4}}\hskip-6.9pt%
        \raisebox{3pt}{\drawsquare{6.5}{0.4}}\hskip-0.4pt
        \raisebox{3pt}{\drawsquare{6.5}{0.4}}}
\newcommand{\Ysquare}{\raisebox{-3.5pt}{\drawsquare{6.5}{0.4}}\hskip-0.4pt%
        \raisebox{-3.5pt}{\drawsquare{6.5}{0.4}}\hskip-13.4pt%
        \raisebox{3pt}{\drawsquare{6.5}{0.4}}\hskip-0.4pt%
        \raisebox{3pt}{\drawsquare{6.5}{0.4}}}
\newcommand{\Yflavor}{\Yfund + \overline{\Yfund}} 
\newcommand{\Yoneoone}{\raisebox{-3.5pt}{\drawsquare{6.5}{0.4}}\hskip-6.9pt%
        \raisebox{3pt}{\drawsquare{6.5}{0.4}}\hskip-6.9pt%
        \raisebox{9.5pt}{\drawsquare{6.5}{0.4}}\hskip-0.4pt%
        \raisebox{9.5pt}{\drawsquare{6.5}{0.4}}}%

\newcommand{\bra}[1]{\langle #1|}
\newcommand{\get}[1]{| #1\rangle}
\def\tr{\mathrm{Tr}}

\newcommand{\beq}{\begin{equation}}
\newcommand{\eeq}{\end{equation}}

\newcommand{\be}{\begin{eqnarray}}
\newcommand{\ee}{\end{eqnarray}}
\title{Technicolor and Beyond: 
Unification in Theory Space}

\author{Francesco Sannino}

\address{CP$^3$-Origins, University of Southern Denmark, Odense M. Denmark. 
\\{ \tiny CP$^3$-Origins-2010-44}
}

\ead{sannino@cp3.sdu.dk, sannino@cp3-origins.net}

\begin{abstract}
The salient features of models of dynamical electroweak symmetry breaking are reviewed. The {\it ideal walking} idea is introduced according to which one should carefully take into account the effects of the extended technicolor dynamics on the technicolor dynamics itself. The effects amount at the enhancement of the anomalous dimension of the mass of the techniquarks  allowing to decouple the Flavor Changing Neutral Currents problem from the one of the generation of the top mass. 

Precision data constraints are reviewed focussing on the latest crucial observation that the $S$-parameter can be computed exactly near the upper end of the conformal window ({\it Conformal S-parameter}) with relevant consequences on the selection of nature's next strong force. We will then introduce the  Minimal Walking Technicolor (MWT) models.

In the second part of this review we consider the interesting possibility to marry supersymmetry and technicolor. The reason is to provide a unification of different extensions of the standard model. {}For example, this means that one can recover, according to the parameters and spectrum of the theory distinct extensions of the standard model, from supersymmetry to technicolor and unparticle physiscs. A surprising result is that a minimal (in terms of the smallest number of fields) supersymmetrization of the MWT model leads to the maximal supersymmetry in four dimensions, i.e. ${\cal N}=4$ SYM. 
\end{abstract}

\section{The Need to Go Beyond}

The standard model (SM) of particle interactions passes a large number of experimental tests. Yet we know that it cannot be the ultimate model of nature since it fails to explain the origin of the matter-antimatter asymmetry and does not provide phenomenologically viable dark matter candidates.  Several extensions of the SM have been proposed, and two stand out in the quest of a better theory: Supersymmetry and Technicolor (TC). We will review here the basics of TC, the problems intrinsically related to the early models of extended TC (ETC) interactions and propose  some concrete solutions. We, in fact, suggest that the best models of TC having a chance to solve the top mass problem together with the suppression of the Flavor Changing Neutral Currents problem, must take into account the effects of the four-fermion interactions stemming from the unspecified ETC dynamics on the TC dynamics. We will argue that when this effect is taken properly into account it will lead to  {\it ideal walking} models which are phenomenologically viable. 

In the second part of this review  we will adopt a more  {\it liberal} attitude towards possible extensions of the SM while still embracing the view that the new extension of the SM should be natural.  We will introduce the new concept of {\it Unification in Theory Space}, according to which our new model will feature supersymmetry and a new gauge theory coupled to the SM a l\'a TC, but also supersymmetric.

\begin{figure}
\includegraphics[width=15cm]{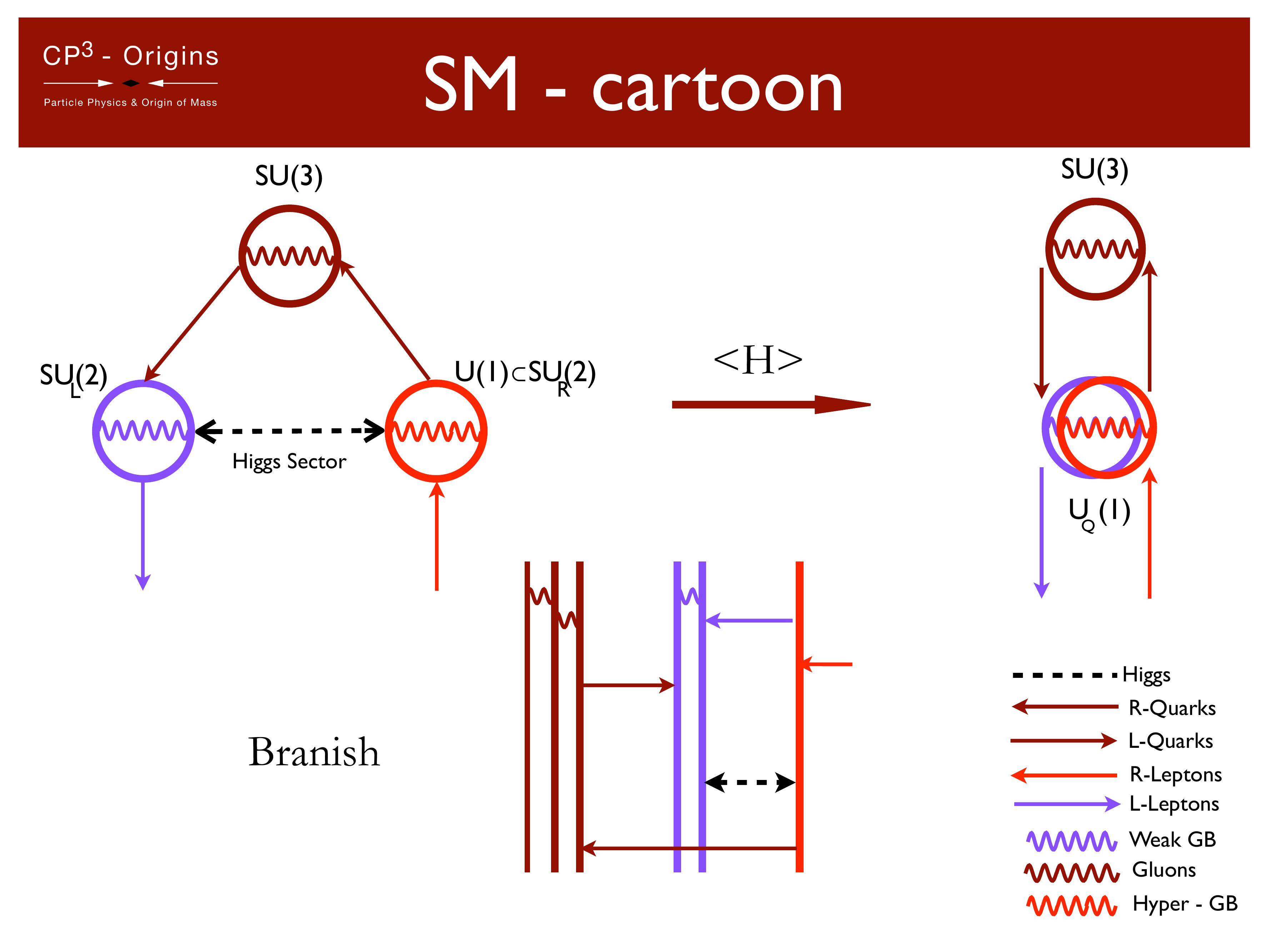}
\label{SM-cartoon}
\caption{Graphical representation of the SM and the Higgs mechanism.  The solid lines connecting the different blobs (gauge groups) represent the SM fermions while the dashed line is the SM Higgs. The wavy-lines inside the blobs are the SM gauge bosons.
}
\end{figure}

We start by elucidating graphically the SM by considering each of the gauge groups independently in Fig.~{\ref{SM-cartoon}}. The solid lines connecting the different blobs (gauge groups) represent the SM fermions while the dashed line is the SM Higgs. The wavy-lines inside the blobs are the SM gauge bosons. 
In the same figure the phenomenon of spontaneous symmetry breaking is graphically represented by shortcutting the Higgs dashed line causing the $SU_L(2)$ and $U(1)$ hypercharge blobs  to superimpose. At the bottom of the figure the blobs are replaced by straight lines naively representing stacks of D3-branes from a possible stringy origin of the SM.

\section{Dynamical Electroweak Symmetry Breaking}

Even in complete absence of the Higgs sector in the standard model (SM) the electroweak symmetry breaks \cite{Farhi:1980xs} due to the condensation of the following quark bilinear in QCD: 
\beq \langle\bar u_Lu_R + \bar d_Ld_R\rangle \neq 0 \ . \label{qcd-condensate}\eeq
and its contribution to the $W$-gauge boson mass is: 
\beq M_W = \frac{gF_\pi}{2} \sim 29 {\rm MeV}\ , \eeq 
with $F_{\pi}\simeq 93$~MeV the pion decay constant. This contribution is, of course, too small when compared to the actual value of the $M_W$ mass that one typically neglects it.   

According to the original idea of technicolor \cite{Weinberg:1979bn,Susskind:1978ms} one augments the SM with another gauge interaction similar to QCD but with a new dynamical scale of the order of the electroweak one. It is sufficient that the new gauge theory is asymptotically free and has global symmetry able to contain the SM $SU_L(2)\times U_Y(1)$ symmetries. It is also required that the new global symmetries break dynamically in such a way that the embedded $SU_L(2)\times U_Y(1)$  breaks to the electromagnetic abelian charge $U_{Q}(1)$. The dynamically generated scale will then be fit to the electroweak one. Note that, except in certain cases, dynamical behaviors are typically nonuniversal which means that different gauge groups and/or matter representations will, in general, posses different dynamics.

The simplest example of technicolor theory is the scaled up version of QCD, i.e. an $SU(N_{TC})$  nonabelian gauge theory with two Dirac Fermions transforming according to the fundamental representation or the gauge group. We need at least two Dirac flavors  to realize the $SU_L(2) \times SU_R(2)$ symmetry of the SM discussed in the SM Higgs section. One simply chooses the scale of the theory to be such that the new pion decaying constant is: \beq F_\pi^{TC} = v_{\rm weak} \simeq 246~ {\rm GeV} \ . \eeq This model correctly accounts for the electroweak symmetry breaking but unfortunately these first attempts are disfavored by precision data.

We have recently proposed several minimal working models \cite{Sannino:2004qp, Hong:2004td,Dietrich:2005wk,Dietrich:2005jn,Gudnason:2006mk,Ryttov:2008xe,Frandsen:2009fs,Frandsen:2009mi,Antipin:2009ks} passing precision data and possessing interesting dynamics relevant for collider phenomenology \cite{Foadi:2007ue,Belyaev:2008yj,Antola:2009wq,Antipin:2010it} and cosmology \cite{Nussinov:1985xr,Barr:1990ca,Bagnasco:1993st,Gudnason:2006ug,Gudnason:2006yj,Kainulainen:2006wq,Kouvaris:2007iq,Kouvaris:2007ay,Khlopov:2007ic,Khlopov:2008ty,Kouvaris:2008hc,Belotsky:2008vh,Cline:2008hr,Nardi:2008ix,Foadi:2008qv,Jarvinen:2009wr,Frandsen:2009mi,Jarvinen:2009mh,Kainulainen:2009rb,Kainulainen:2010pk,Frandsen:2010yj,Kouvaris:2010vv} \footnote{Note that if dark matter arises as a pseudo Goldstone boson in technicolor models (the TIMP) \cite{Gudnason:2006ug,Gudnason:2006yj,Nardi:2008ix,Foadi:2008qv,Jarvinen:2009wr,Frandsen:2009mi,Jarvinen:2009mh} they should not be confused  with the scaled up version of the ordinary baryon envisioned earlier \cite{Nussinov:1985xr,Barr:1990ca,Bagnasco:1993st}. {}For example one striking feature is that the pseudo Goldstone Bosons can be sufficiently light to be produced at the Large Hadron Collider experiment at CERN.}. 

These models are also being investigated via first principle lattice simulations \cite{Catterall:2007yx,Catterall:2008qk,DelDebbio:2008zf,Hietanen:2008vc,Hietanen:2009az,Pica:2009hc,Catterall:2009sb,Lucini:2009an,Bursa:2009we,DelDebbio:2010hu,DelDebbio:2010hx,DeGrand:2009hu,DeGrand:2008kx,DeGrand:2009mt,Fodor:2008hm,Fodor:2009ar,Fodor:2009nh,Kogut:2010cz}. An up-to-date review is Ref. \cite{Sannino:2009za,Sannino:2008ha}. 

\begin{figure}
\includegraphics[width=15cm]{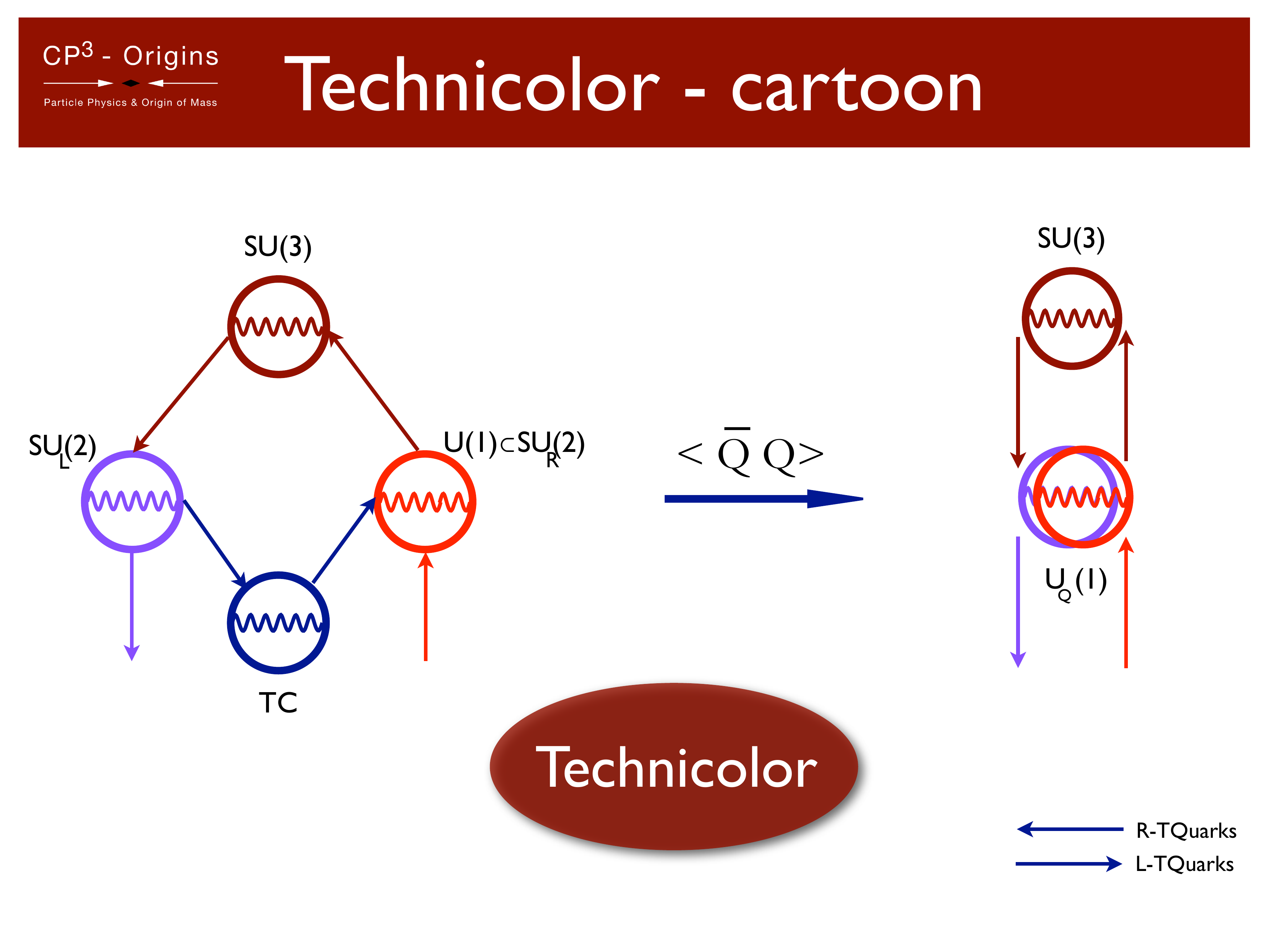}
\label{TC-cartoon}
\caption{Graphical representation of the SM and the Higgs mechanism replaced by a new TC gauge dynamics. }
\end{figure}

A simple graphical representation of TC is shown in Fig.~\ref{TC-cartoon} in which the SM Higgs (dashed line in  Fig.~\ref{SM-cartoon}) is replaced by a new gauge dynamics. 
We do not show, in both figures, the fact that the color interactions contribute to the shortcutting of the electroweak interactions given that their contribution can be neglected, in first approximation, with respect to the one due to either the SM Higgs mechanism or the TC one. 
\subsection{Constraints from Electroweak Precision Data}
\label{5}

\begin{figure}[b]
\begin{center}
\includegraphics[width=9truecm,height=9truecm]{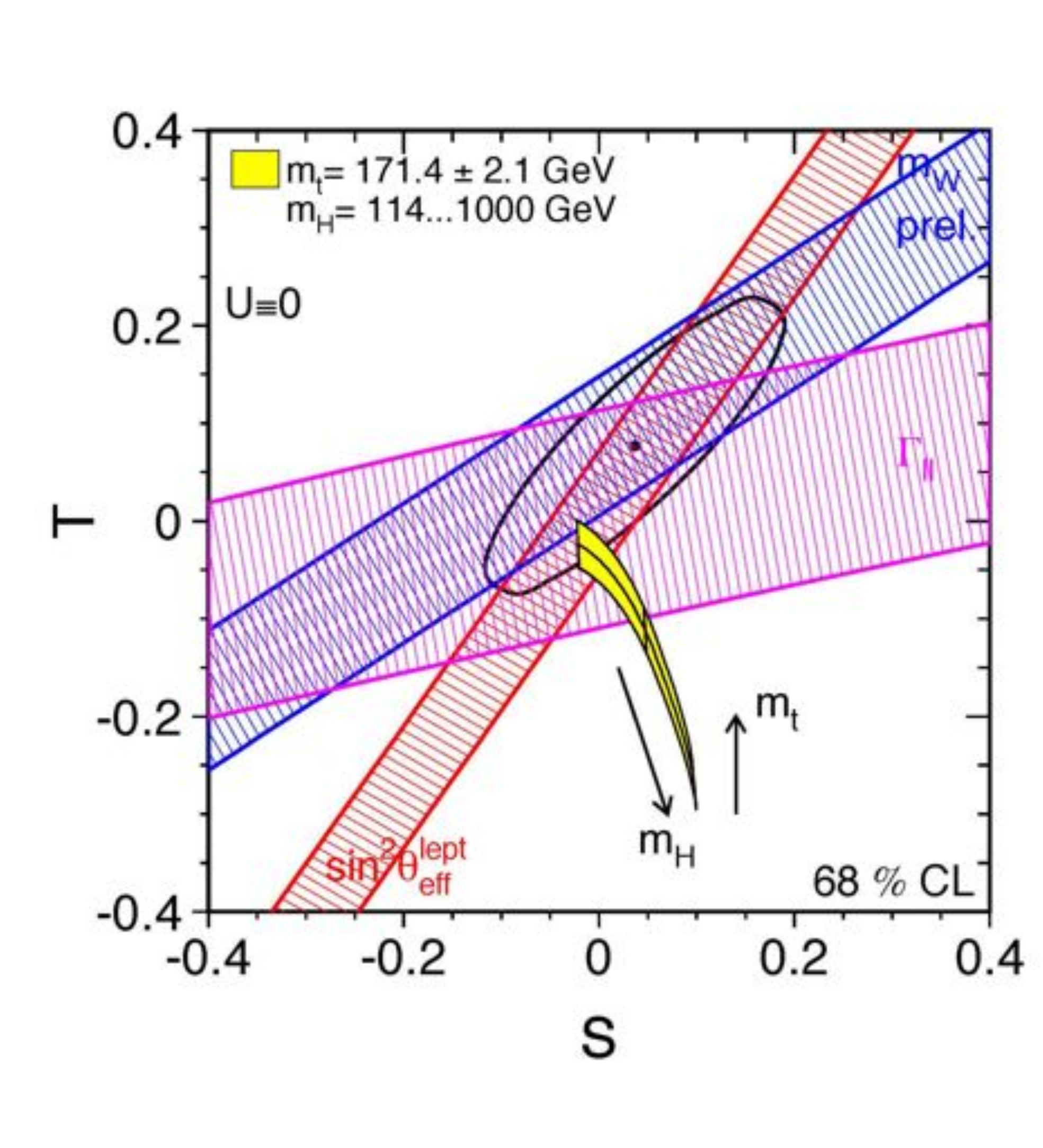}
\caption{The 1-$\sigma$ range of the electroweak parameters $S$ and $T$ determined from different observables. The ellipsis shows the 68\% probability from combined data. The yellow area gives the SM prediction with $m_t$ and $m_H$ varied as shown.} \label{s06_stu_contours}
\end{center}
\end{figure}

The relevant corrections due to the presence of new physics trying to modify the electroweak breaking sector of the SM appear in the vacuum polarizations of the electroweak gauge bosons. These can be parameterized in terms of the three 
quantities $S$, $T$, and $U$ (the oblique parameters) 
\cite{Peskin:1990zt,Peskin:1991sw,Kennedy:1990ib,Altarelli:1990zd}, and confronted with the electroweak precision data. Recently, due to the increase precision of the measurements reported by LEP II, the list of interesting parameters to compute has been extended \cite{hep-ph/9306267,Barbieri:2004qk}.  We show below also the relation with the traditional one \cite{Peskin:1990zt}.  Defining with  $Q^2\equiv -q^2$ the Euclidean transferred momentum entering in a generic two point function vacuum polarization associated to the electroweak gauge bosons, and denoting derivatives with respect to $-Q^2$ with a prime we have \cite{Barbieri:2004qk}: 
\begin{eqnarray}
\hat{S} &\equiv & g^2 \ \Pi_{W^3B}^\prime (0) \ , \\
\hat{T} &\equiv & \frac{g^2}{M_W^2}\left[ \Pi_{W^3W^3}(0) -
\Pi_{W^+W^-}(0) \right] \ , \\
W &\equiv & \frac{g^2M_W^2}{2} \left[\Pi^{\prime\prime}_{W^3W^3}(0)\right] \ , \\
Y &\equiv & \frac{g'^2M_W^2}{2} \left[\Pi^{\prime\prime}_{BB}(0)\right] \ , \\
\hat{U} &\equiv & -g^2 \left[\Pi^\prime_{W^3W^3}(0)-
\Pi^\prime_{W^+W^-}(0)\right]\ , \\
V &\equiv & \frac{g^2 \, M^2_W}{2}\left[\Pi^{\prime\prime}_{W^3W^3}(0)-
\Pi^{\prime\prime}_{W^+W^-}(0)\right] \ , \\
X &\equiv & \frac{g g'\,M_W^2}{2} \ \Pi_{W^3B}^{\prime\prime}(0) \ .
\end{eqnarray}
Here $\Pi_V(Q^2)$ with $V=\{W^3B,\, W^3W^3,\, W^+W^-,\, BB\}$ represents the
self-energy of the vector bosons. Here the
electroweak couplings are the ones associated to the physical electroweak gauge bosons:
\begin{eqnarray}
\frac{1}{g^2} \equiv  \Pi^\prime_{W^+W^-}(0)
 \ , \qquad \frac{1}{g'^2}
\equiv  \Pi^\prime_{BB}(0) \ ,
\end{eqnarray}
while $G_F$ is
\begin{eqnarray}
\frac{1}{\sqrt{2}G_F}=-4\Pi_{W^+W^-}(0) \ ,
\end{eqnarray}
as in \cite{Chivukula:2004af}. $\hat{S}$ and $\hat{T}$ lend their name
from the well known Peskin-Takeuchi parameters $S$ and $T$ which are related to the new ones via
\cite{Barbieri:2004qk,Chivukula:2004af}:
\begin{eqnarray}
\frac{\alpha S}{4s_W^2} =  \hat{S} - Y - W  \ , \qquad 
\alpha T = \hat{T}- \frac{s_W^2}{1-s_W^2}Y \ .
\end{eqnarray}
Here $\alpha$ is the electromagnetic structure constant and $s_W=\sin \theta_W $
is the weak mixing angle. Therefore in the case where $W=Y=0$ we
have the simple relation
\begin{eqnarray}
\hat{S} &=& \frac{\alpha S}{4s_W^2} \ , \qquad 
\hat{T}= \alpha T \ .
\end{eqnarray}
The result of the the fit is shown in Fig.~\ref{s06_stu_contours}. If the value of the Higgs mass increases the central value of the $S$ parameters moves to the left towards negative values. 

\noindent
In technicolor it is easy to have a vanishing $T$ parameter while typically $S$ is positive. Besides, the composite Higgs is typically heavy with respect to the Fermi scale, at least for technifermions in the fundamental representation of the gauge group and for a small number of techniflavors. The oldest technicolor models featuring QCD dynamics with three technicolors and a doublet of electroweak gauged techniflavors deviate a few sigmas from the current precision tests.

Clearly it is desirable to reduce the tension between the precision data and a possible dynamical mechanism underlying the electroweak symmetry breaking. It is possible to imagine different ways to achieve this goal and some of the earlier attempts have been summarized in \cite{Peskin:2001rw}. 

The computation of the $S$ parameter in technicolor theories requires the knowledge of nonperturbative dynamics rendering difficult the precise knowledge of the contribution to  $S$. {}For example, it is not clear what is the exact value of the composite Higgs mass relative to the Fermi scale and, to be on the safe side, one typically takes it to be quite large, of the order at least of the TeV. However in certain models it may be substantially lighter due to the intrinsic dynamics \cite{Hong:2004td,Dietrich:2005wk,Dietrich:2005jn}. We will discuss the spectrum of different strongly coupled theories in the Appendix and its relation to the electroweak parameters later in this chapter.

 It is, however, instructive to provide a simple estimate of the contribution to $S$ which allows to guide model builders. Consider a one-loop exchange of $N_D$ doublets of techniquarks transforming according to the representation $R_{TC}$ of the underlying technicolor gauge theory and with dynamically generated mass $\Sigma_{(0)}$ assumed to be larger than the weak intermediate gauge bosons masses. Indicating with $d(R_{\rm TC})$ the dimension of the techniquark representation, and to leading order in $M_{W}/\Sigma(0)$ one finds:
 \begin{eqnarray}
S_{\rm naive} = N_D \frac{d(R_{\rm TC})}{6\pi} \ .
\end{eqnarray} 
This naive value provides, in general, only a rough estimate of the exact value of $S$. 
However, it is clear from the formula above that, the more technicolor matter is gauged under the electroweak theory the larger is the $S$ parameter and that the final $S$ parameter is expected to be positive. We will show that this is, indeed, the case as it can be argued by investigating the same correlator in regimes of the underlying gauge theory where it becomes exactly calculable.  

Attention must be paid to the fact that the phenomenologically relevant $S$ parameter can receive contributions also from other sectors. Such a contribution can be taken sufficiently large and negative to compensate for the positive value from the composite Higgs dynamics. To be concrete: Consider an extension of the SM in which the Higgs is composite but we also have new heavy (with a mass of the order of the electroweak) fourth family of Dirac leptons. In this case a sufficiently large splitting of the new lepton masses can strongly reduce and even offset the positive value of $S$.  The total $S$ can be written as:
\begin{eqnarray}
S = S_{\rm TC} + S_{\rm NS} \ .
\end{eqnarray}
 The parameter $T$ will be, in general, modified and one has to make sure that the corrections do not spoil the agreement with this parameter.  From the discussion above it is clear that technicolor models can be constrained, via precision measurements, only model by model and the effects of possible new sectors must be properly included
 
The presentation so far is incomplete. In fact it neglects the constraints and back-reaction on the gauge sector coming from the one giving masses to the SM fermions. To estimate these effects we have considered two simple extensions able, in an effective way, to accommodate the SM masses. In \cite{Fukano:2009zm}, to estimate these corrections,  the composite Higgs sector was coupled directly to the SM fermions \cite{Foadi:2007ue}. Here one gets relevant constraints on the $W$ parameter while the corrections do not affect the $S$-parameter. The situation changes when an entirely new sector is introduced in the flavor sector. Due to the almost inevitable interplay between the gauge and the flavor sector the back-reaction of the flavor sector is very relevant \cite{Antola:2009wq}.  Mimicking the new sector via a new (composite or not) Higgs coupling directly to the SM fermions it was observed that important corrections to the $S$ and $T$ parameters arise which can be used to compensate a possible heavy composite Higgs scenario of the technicolor sector \cite{Antola:2009wq}. To investigate these effects we adopted a straightforward and instructive model according to which we have both a composite sector and  a fundamental scalar field  (SM-like Higgs) intertwined at the electroweak scale. This idea was pioneered  in a series of papers by Simmons  \cite{Simmons:1988fu},  Dine,  Kagan and Samuel \cite{Dine:1990jd,Samuel:1990dq,Kagan:1991gh,Kagan:1992aq,Kagan:1990gi} and Carone and Georgi  \cite{Carone:1992rh,Carone:1994mx}. More recently this type of model has been investigated also in \cite{Hemmige:2001vq,Carone:2006wj,Zerwekh:2009yu}. Interesting related work can be also found in \cite{Chivukula:1990bc,Chivukula:2009ck}.

\subsection{Standard Model Fermion Masses}

Since in a purely technicolor model  the Higgs is a composite particle the Yukawa terms, when written in terms of the underlying technicolor fields, amount to four-fermion operators. The latter can be naturally interpreted as low energy operators induced by a new strongly coupled gauge interaction emerging at energies higher than the electroweak theory. These type of theories have been termed extended technicolor interactions (ETC) \cite{Eichten:1979ah,Dimopoulos:1979es}. 

Here we describe the simplest ETC model in which the ETC interactions connect the chiral symmetries of the techniquarks to those of the SM fermions.

Let's start with the case in which the ETC dynamics is represented by a $SU(N_{ETC})$ gauge group with: 
\beq N_{ETC} = N_{TC} + N_g \ , \eeq
and $N_g$ is the number of SM
generations. In order to give masses to all of the SM fermions, in this scheme, one needs a condensate for each SM fermion. This can be achieved by using as technifermion matter a complete generation of quarks and leptons (including a neutrino right) but now gauged with respect to the technicolor interactions.  

The ETC gauge group is assumed to spontaneously break $N_g$ times down to
$SU(N_{TC})$ permitting
three different mass scales, one  for each SM family. This type of technicolor with associated ETC is
termed the \emph{one family model} \cite{Farhi:1979zx}.
The heavy masses are provided by the
breaking at low energy and the light masses are provided by breaking
at higher energy scales. 
This model does not, per se, explain how the
gauge group is broken several times, neither is the breaking of weak isospin
symmetry accounted for.

Schematically one has $SU(N_{TC} + 3)$ which breaks to  $SU(N_{TC} + 2)$ at the scale 
$\Lambda_1$ providing the first generation of fermions with a typical mass $m_1 \sim {4\pi
  (F_\pi^{TC})^3}/{\Lambda_1^2}$ at this point the gauge group breaks to $SU(N_{TC} + 1)$ with dynamical scale $\Lambda_2 $ leading to a second generation mass of the order of $m_2 \sim{4\pi
  (F_\pi^{TC})^3}/{\Lambda_2^2}$ finally the last breaking
$SU(N_{TC} )$ at scale 
$\Lambda_3$ leading to the last generation mass $m_3 \sim {4\pi
  (F_\pi^{TC})^3}/{\Lambda_3^2}$. 
 
\begin{figure}
\begin{center}
\includegraphics[width=16cm]{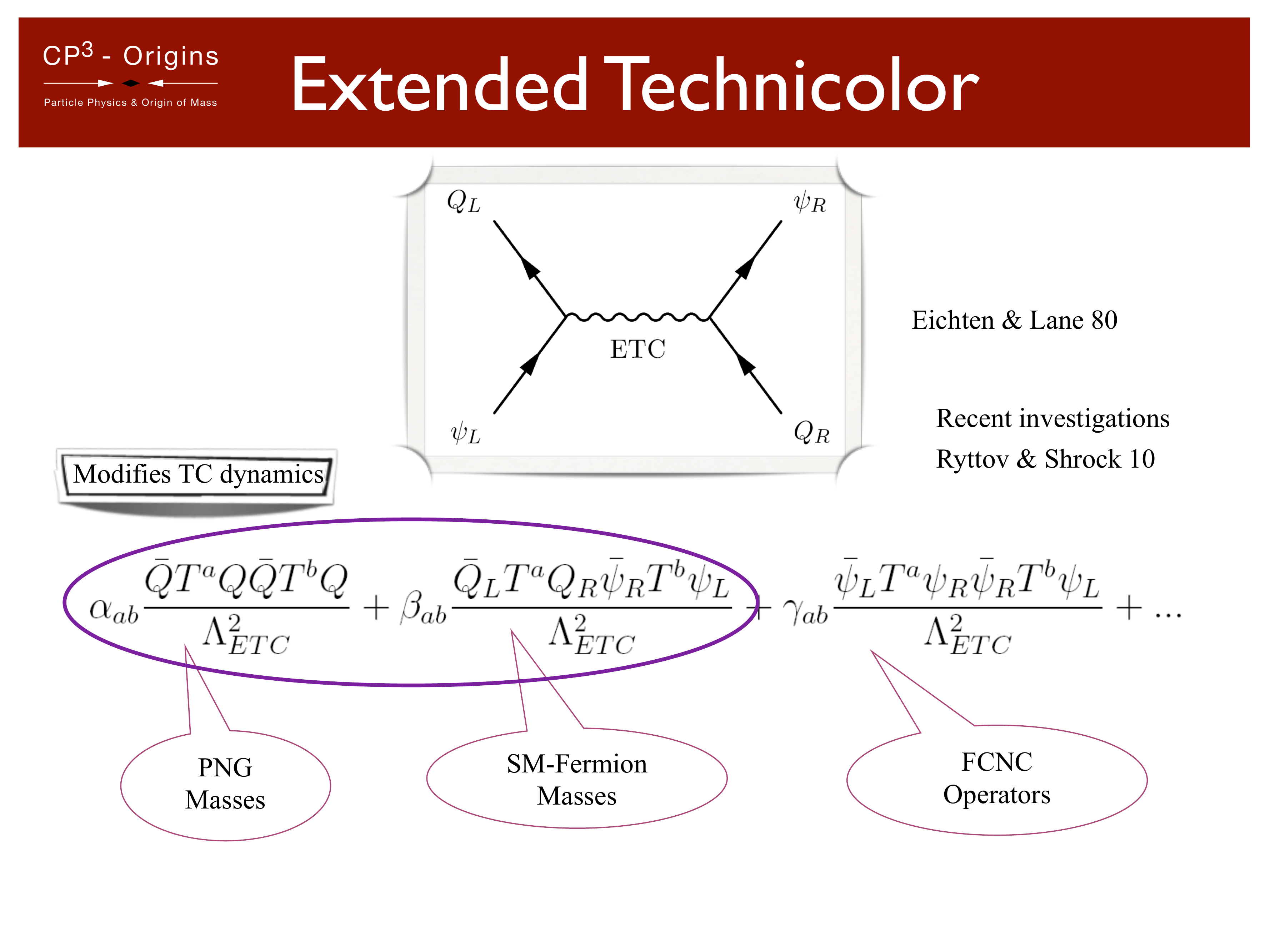}
\end{center}
\caption{ETC schematic elementary interactions and the expected low energy four-fermion interactions. }
\label{ETC1}
\end{figure}

Without specifying an ETC one can write down the most general type of four-fermion operators involving technicolor particles $Q$ and ordinary fermionic fields $\psi$.  {}Following the notation of Hill and Simmons \cite{Hill:2002ap} we write:
\beq \alpha_{ab}\frac{\bar Q\gamma_\mu T^aQ\bar\psi \gamma^\mu
  T^b\psi}{\Lambda_{ETC}^2} +
\beta_{ab}\frac{\bar Q\gamma_\mu T^aQ\bar Q\gamma^\mu
  T^bQ}{\Lambda_{ETC}^2} + 
\gamma_{ab}\frac{\bar\psi\gamma_\mu T^a\psi\bar\psi\gamma^\mu
  T^b\psi}{\Lambda_{ETC}^2} \ , \eeq
where the $T$s are unspecified ETC generators. After performing a Fierz rearrangement one has:
\beq \alpha_{ab}\frac{\bar QT^aQ\bar\psi T^b\psi}{\Lambda_{ETC}^2} +
\beta_{ab}\frac{\bar QT^aQ\bar QT^bQ}{\Lambda_{ETC}^2} +
\gamma_{ab}\frac{\bar\psi T^a\psi\bar\psi T^b\psi}{\Lambda_{ETC}^2}
+ \ldots \ , \label{etc} \eeq
The coefficients parametrize the ignorance on the specific ETC physics. We summarize in Fig.~\ref{ETC1} the ETC schematic elementary interactions and the expected low energy four-fermion interactions.
To be more specific, the $\alpha$-terms, after the technicolor particles have condensed, lead to mass terms for the SM fermions
\beq m_q \approx \frac{g_{ETC}^2}{M_{ETC}^2}\langle \bar
QQ\rangle_{ETC} \ , \eeq
where $m_q$ is the mass of e.g.~a SM quark, $g_{ETC}$ is the ETC gauge 
coupling constant evaluated at the ETC scale, $M_{ETC}$ is the mass of
an ETC gauge boson and $\langle \bar QQ\rangle_{ETC}$ is the
technicolor condensate where the operator is evaluated at the ETC
scale. Note that we have not explicitly considered the different scales for the different generations of ordinary fermions but this should be taken into account for any realistic model. 

The $\beta$-terms of Eq.~(\ref{etc}) provide masses for
pseudo Goldstone bosons and also provide masses for techniaxions
\cite{Hill:2002ap}. 
The last class of terms, namely the $\gamma$-terms of
Eq.~(\ref{etc}) induce flavor changing neutral currents. {}For example it may generate the following terms:
\beq \frac{1}{\Lambda_{ETC}^2}(\bar s\gamma^5d)(\bar s\gamma^5d) +
\frac{1}{\Lambda_{ETC}^2}(\bar \mu\gamma^5e)(\bar e\gamma^5e) + 
\ldots \ , \label{FCNC} \eeq
where $s,d,\mu,e$ denote the strange and down quark, the muon
and the electron, respectively. The first term is a $\Delta S=2$
flavor-changing neutral current interaction affecting the
$K_L-K_S$ mass difference which is measured accurately. The experimental bounds on these type of operators together with the very {\it naive} assumption that ETC will generate these operators with $\gamma$ of order one leads to a constraint on the ETC scale to be of the order of or larger than $10^3$
TeV \cite{Eichten:1979ah}. This should be the lightest ETC scale which in turn puts an upper limit on how large the ordinary fermionic masses can be. The naive estimate is that  one can account up to around 100 MeV mass for a QCD-like technicolor theory, implying that the top quark mass value cannot be achieved. 

The second term of Eq.~(\ref{FCNC}) induces flavor
changing processes in the leptonic sector such as $\mu\rightarrow e\bar ee,
e\gamma$ which are not observed.
It is clear that, both for the precision measurements  and the fermion masses, that a better theory of the flavor is needed. {}For the ETC dynamics interesting  developments recently appeared in the literature  \cite{Ryttov:2010fu,Ryttov:2010hs,Ryttov:2010jt,Ryttov:2010kc}. We note that nonperturbative chiral gauge theories dynamics is expected to play a relevant role in models of ETC since it allows, at least in principle, the self breaking of the gauge symmetry. Recent progress on the phase diagrams of these theories has appeared in \cite{Sannino:2009za}.  

\begin{figure}
\begin{center}
\includegraphics[width=16truecm]{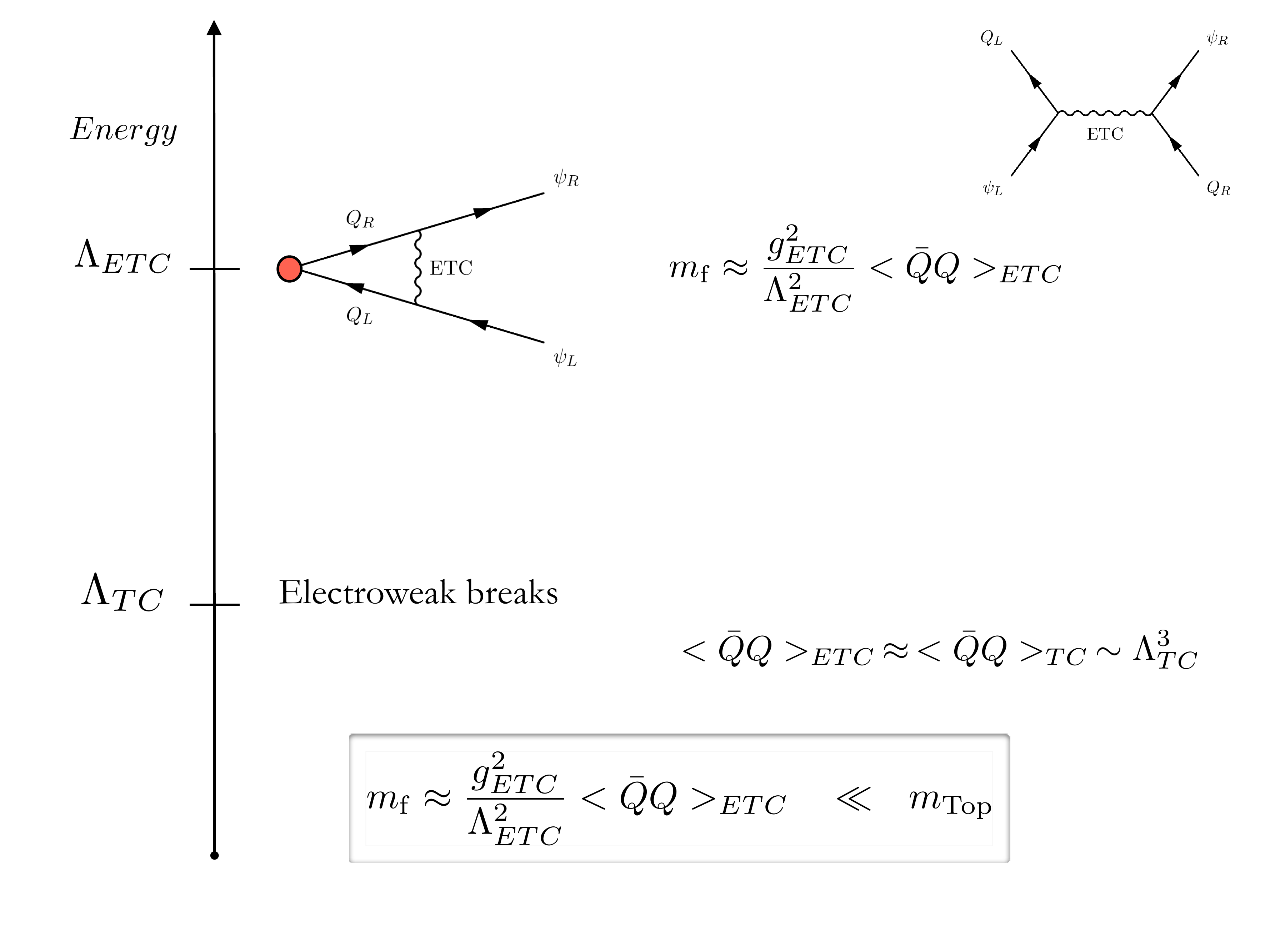}
\end{center}
\caption{Cartoon of the expected ETC dynamics starting at high energies with a more fundamental gauge interaction and the generation of the fermion masses assuming QCD-like dynamics.}
\label{ETC-dyn}
\end{figure}

In Figure \ref{ETC-dyn} we show the ordering of the relevant scales involved in the generation of the ordinary fermion masses via ETC dynamics, and the generation of the fermion masses (for a single generation and focussing on the top quark)  assuming QCD-like dynamics for TC.

\subsection{Ideal Walking}
To understand how to modify the QCD dynamics we analyze the TC condensate. 
The value of the technicolor condensate used when giving mass to the ordinary fermions should be evaluated not at the technicolor scale but at the extended technicolor one. Via the renormalization group one can relate the condensate at the two scales via:
\beq \langle\bar QQ\rangle_{ETC} =
\exp\left(\int_{\Lambda_{TC}}^{\Lambda_{ETC}}
d(\ln\mu)\gamma_m(\alpha(\mu))\right)\langle\bar QQ\rangle_{TC} \ ,
\label{rad-cor-tc-cond}
\eeq 
where $\gamma_m$ is the anomalous dimension of the techniquark mass-operator. The boundaries of the integral 
 are at the ETC scale and the TC one.
{}For TC theories with a running of the coupling constant similar to the one in QCD, i.e.
\beq \alpha(\mu) \propto \frac{1}{\ln\mu} \ , \quad {\rm for}\ \mu >
\Lambda_{TC} \ , \eeq
this implies that the anomalous dimension of the techniquark masses $\gamma_m \propto
\alpha(\mu)$. When computing the integral one gets
\beq \langle\bar QQ\rangle_{ETC} \sim
\ln\left(\frac{\Lambda_{ETC}}{\Lambda_{TC}}\right)^{\gamma_m}
\langle\bar QQ\rangle_{TC} \ , \label{QCD-like-enh} \eeq
which is a logarithmic enhancement of the operator. We can hence neglect this correction and use directly the value of the condensate at the TC scale when estimating the generated fermionic mass:
\beq m_q \approx \frac{g_{ETC}^2}{M_{ETC}^2}\Lambda_{TC}^3 \ , \qquad 
 \langle \bar QQ\rangle_{TC} \sim \Lambda_{TC}^3 \ . \eeq

The tension between having to reduce the FCNCs and at the same time provide a sufficiently large mass for the heavy fermions in the SM as well as the pseudo-Goldstones can be reduced if the dynamics of the underlying TC theory is different from the one of QCD. The computation of the TC condensate, at different scales, shows that  if the dynamics is such that the TC coupling does not {\it run} to the UV fixed point but rather slowly reduces to zero one achieves a net enhancement of the condensate itself with respect to the value estimated earlier.  This can be achieved if the theory has a near conformal fixed point. This kind of dynamics has been denoted of {\it walking} type. In this case 
\beq \langle\bar QQ\rangle_{ETC} \sim
\left(\frac{\Lambda_{ETC}}{\Lambda_{TC}}\right)^{\gamma_m(\alpha^*)}
\langle\bar QQ\rangle_{TC} \ , \label{walking-enh} \eeq
which is a much larger contribution than in QCD dynamics  \cite{Yamawaki:1985zg,Holdom:1984sk,Holdom:1981rm,Appelquist:1986an}. Here $\gamma_m$ is evaluated at the would be fixed point value $\alpha^*$.  Walking can help resolving the problem of FCNCs in
technicolor models since with a large enhancement of the $\langle\bar
QQ\rangle$ condensate the four-fermi operators involving SM fermions and
technifermions and the ones involving technifermions are enhanced by a factor of
$\Lambda_{ETC}/\Lambda_{TC}$ to the $\gamma_m$  power while the one involving only SM fermions is not enhanced.

There are several issues associated with the original idea of walking: 
\begin{itemize}
\item{Since the number of flavors cannot be changed continuously it is not possible to  get arbitrarily close to the lower end of the conformal window. This applies to the technicolor theory {\it in isolation} i.e. before coupling it to the SM and without taking into account the ETC interactions. }
\item{It is hard to achieve large anomalous dimensions of the fermion mass operator even near the lower end of the conformal window for ordinary gauge theories.}
\item{It is not always possible to neglect the interplay of the four fermion interactions on the TC dynamics. }
\end{itemize}

We have  argued in \cite{Fukano:2010yv} that it is possible to {\it solve} simultaneously all the problems above by consistently taking into account the effects of the four-fermion interactions on the phase diagram of strongly interacting theories for any matter representation as function of the number of colors and flavors.  A positive effect is that the anomalous dimension of the mass increases beyond the unity value at the lower boundary of the new conformal window and can get sufficiently large to yield the correct mass for the top quark. We have also shown that the conformal window, for any representation, shrinks with respect to the case in which the four-fermion interactions are neglected. 

We made the further {\it unexpected} discovery that when the extended TC sector, responsible for giving masses to the standard model fermions, is sufficiently strongly coupled the TC theory, in isolation,  must  feature an infrared fixed point in order for the full model to be phenomenologically viable and correctly break the electroweak symmetry.

 Using the new phase diagram we showed that the minimal walking TC models are the archetypes of models of dynamical electroweak symmetry breaking. 

The presence of the four-fermion interactions allows to get sufficiently near the lower end of the conformal window using as continuous variable the strength of the four-fermion interactions.

\section{Conformal S-Parameter, Exact results,  Lower Bound Conjecture and Gauge Dualities.}
It was noted, for the first time in \cite{Sannino:2010ca}, that the $VV-AA$ correlator can be computed exactly, in perturbation theory, near the upper end of the conformal window of any gauge theory of fundamental interaction.  A mass term operator  was also introduced  to {\it probe} the dynamics of the theory. Note that adding relevant operators is a standard procedure in field theory and statistical mechanics when investigating the properties associated to any exact symmetry or critical phenomena respectively. The resulting correlator was termed {\it Conformal S-Parameter} and it was also argued that it is smoothly connected to the physically relevant $S$-parameter below the lower end of the conformal window \cite{Sannino:2010ca} allowing for important phenomenological consequences such as: {\it one-family technicolor models are ruled out by precision data even if they would be walking}. This statement does not extend to minimal technicolor models constructed via, for example, higher dimensional representations with respect to the underlying technicolor gauge group. 

The one-loop results were generalized in \cite{DiChiara:2010xb} by taking into account the  2-loop corrections at the upper end of the conformal window. We arrived at the following important formula for the Conformal $S$-parameter \cite{DiChiara:2010xb} :
\be
\lim_{\frac{q^2}{m^2}\rightarrow 0} \frac{6\pi S}{\sharp } = 1 +\frac{17}{72}\gamma_m(\alpha^*)\, ,
\label{Sgamma}
\ee
with
\be
\gamma_m(\alpha)=\frac32 C_2\left[r\right] \frac{\alpha}{\pi}\, ,
\ee
 $\alpha^{\ast}$ the two-loops fixed point value of the gauge coupling, $C_2\left[r\right]$ the quadratic casimir of the representation $r$,  $\gamma_m$ the mass term anomalous dimension and $\sharp = N_D \, d[r] $ is the number of doublets $N_D$ times the dimension of the representation $d[r]$ under which the fermions transform and $q^2$ the external momentum.

The above expressions unequivocally show that the normalized $S$-parameter is a decreasing function of $N_f$ near the upper boundary of the conformal window. This important result is in agreement with the conjecture formulated in \cite{Sannino:2010ca}.

\subsection{Magnetic $S$-parameter: The link to gauge duality}

As we decrease the number of flavors, within the conformal window, we have shown that
the normalized $S$ is increasing as we decrease the number of flavors. This statement is exact in perturbation theory and lends further support to the claim made in \cite{Sannino:2010ca} according to which the unity value of the normalized $S$-parameter constitutes the absolute lower bound across the entire phase diagram. 

In formulae the $S$-parameter satisfies:
 \begin{equation}
S_{\rm norm} \equiv \frac{6\pi S}{\sharp} \geq 1  \quad {\rm when} \quad {\frac{q^2}{m^2} \rightarrow 0}  \ .
\label{sbound}
\end{equation}

Beyond perturbation theory it has also been shown \cite{Sannino:2010fh} that near the lower bound of the conformal window the $S$-parameter can be estimated via gauge duality~\cite{Sannino:2009qc,Sannino:2009me,Terning:1997xy}. There is, in fact, the fascinating possibility that generic asymptotically free gauge theories have magnetic duals. These are genuine gauge theories with typically a different gauge group with respect to the original electric theory and matter content. The full theory possesses, however, the same flavor symmetries. At low energy the electric and magnetic theory flow to the same infrared physics. 
The computation of the $S$-parameter would be then possible, in perturbation theory, near the lower bound of the conformal window since the dual gauge theory there is expected to be in a perturbative regime.

 As argued in \cite{Terning:1997xy,Sannino:2009qc,Sannino:2009me} a candidate gauge dual theory to QCD in the conformal window, i.e. an  $SU(3)$ color theory with  a sufficiently large number of massless Dirac flavors ($N_f$), transforming according to the fundamental representation, is constituted by an $SU(X)$ gauge group with global symmetry group $SU_L(N_f)\times SU_R(N_f) \times U_V(1)$  featuring 
{\it magnetic} quarks ${q}$ and $\widetilde{q}$ together with $SU(X)$ gauge singlet fermions identifiable as baryons built out of the {\it electric} quarks $Q$. Since mesons do not affect directly global anomaly matching conditions we can add them to the spectrum of the dual theory. In particular they are needed to let the magnetic quarks and the gauge singlet fermions interact with each others. The new mesons will be massless and have no-self potential to respect the conformal invariance of the model at large distances.  We added to the {\it magnetic} quarks gauge singlet Weyl fermions which can be identified with the baryons of QCD but are, in fact, massless. The generic dual spectrum is summarized in table \ref{dualgeneric}.
\begin{table}[h]
\[ \begin{array}{|c| c|c c c|c|} \hline
{\rm Fields} &\left[ SU(X) \right] & SU_L(N_f) &SU_R(N_f) & U_V(1)& \# ~{\rm  of~copies} \\ \hline 
\hline 
 q &\Yfund &{\Yfund }&1&~~y &1 \\
\widetilde{q} & \overline{\Yfund}&1 &  \overline{\Yfund}& -y&1   \\
A &1&\Ythreea &1&~~~3& \ell_A \\
S &1&\Ythrees &1&~~~3& \ell_S \\
C &1&\Yadjoint &1&~~~3& \ell_C \\
B_A &1&\Yasymm &\Yfund &~~~3& \ell_{B_A} \\
B_S &1&\Ysymm &\Yfund &~~~3& \ell_{B_S} \\
{D}_A &1&{\Yfund} &{\Yasymm } &~~~3& \ell_{{D}_A} \\
{D}_S & 1&{\Yfund}  &{\Ysymm} &  ~~~3& \ell_{{D}_S} \\
\widetilde{A} &1&1&\overline{\Ythreea} &-3&\ell_{\widetilde{A}}\\
\widetilde{S} &1&1&\overline{\Ythrees} & -3& \ell_{\widetilde{S}} \\
\widetilde{C} &1&1&\overline{\Yadjoint} &-3& \ell_{\widetilde{C}} \\
M^i_{j} &1&\Yfund &\overline{\Yfund} & 0 &1 \\
 \hline \end{array} 
\]
\caption{Massless spectrum of {\it magnetic} quarks and baryons and their  transformation properties under the global symmetry group. The last column represents the multiplicity of each state and each state is a  Weyl fermion.}
\label{dualgeneric}
\end{table}

The $\ell$s count the number of times the same baryonic matter representation appears as part of the spectrum of the theory. Invariance under parity and charge conjugation of the underlying theory requires $\ell_{J} = \ell_{\widetilde{J}}$~~ with $J=A,S,...,C$ and $\ell_B = - \ell_D$. 

The simplest mesonic operator is $M_i^{j} $ and transforms simultaneously according to the antifundamental representation of $SU_L(N_f)$ and the fundamental representation of  $SU_R(N_f)$. These states are not constrained by anomaly matching conditions and they mediate the interactions between the magnetic quarks and the gauge singlet fermions via Yukawa-type interactions.  

 Near the lower end of the conformal window the {\it magnetic} S-parameter, i.e. the S-parameter computed in the magnetic theory, is \cite{Sannino:2010fh}: 
 \begin{eqnarray}
S_{m} & = &S_q + S_B +S_{\rm M} \ ,
\end{eqnarray}
with $S_q$, $S_B$ and $S_{\rm M}$ the contributions coming from the magnetic quarks, the baryons and the mesons respectively. 
 In \cite{Sannino:2010fh} it was considered the case in which we gauge, with respect to the electroweak interactions, only the $SU_L (2)\times SU_R (2)$ subgroup where the hypercharge is the diagonal generator of $SU(2)_R$.  In this case only one doublet contributes directly to the $S$-parameter and we have  \cite{Sannino:2010fh}: 
\begin{eqnarray}
\frac{6\pi}{3}S_{m} = \frac{X}{3} + \frac{\ell_C + \ell_{B_A}}{3} + \frac{25}{729} \, \ell_{B_S} \left( 32 \log2 - 39 \right) - 0.14 \ . \nonumber \\
\end{eqnarray} 
 {}From this expression is evident that the present definition of the normalized $S$-parameter {\it counts} the relevant degrees of freedom as function of the number of flavors. We estimated $S_m$  \cite{Sannino:2010fh} using the possible dual provided in \cite{Sannino:2009qc} for which $X = 2 N_f - 15$, $\ell_A = 2$, $\ell_{B_A} = -2 $ (we take $+2$ since we are simply counting the states) with the other $\ell$s vanishing. Asymptotic freedom for the magnetic dual requires at least $N_f=9$ for which $6\pi S_{m}/3 = 1.523$ while if the lower bound of the conformal window occurs for $N_f = 10$ we obtain $6\pi S_{m} /3= 2.19$. Of course, only one of these two values should be considered as the actual value of the normalized magnetic $S$-parameter near the lower end of the electric conformal window. It is quite remarkable that the computation in the magnetic theory in \cite{Sannino:2010fh} yields an estimate which is consistent with the lower bound and the perturbative computations presented here. 

{\subsection{S-parameter Lower Bound:}} 

{\it How can we connect the conformal $S$ with the one below the conformal window?} 

As we decrease the number of flavors we cross into the chirally broken phase and conformality is lost. Below the critical number of flavors corresponding to the lower bound of the conformal window, a dynamical mass of the fermions is generated. In the broken phase we should compute the $S$-parameter, in the zero momentum limit, with the hard mass of the fermions replaced by the hard plus the dynamical one. We noted in \cite{Sannino:2010ca} that this indicates that the broken and symmetric phases are smoothly connected when discussing the normalized $S$-parameter.

\begin{figure}[t]
\begin{center}
\includegraphics[width=10cm]{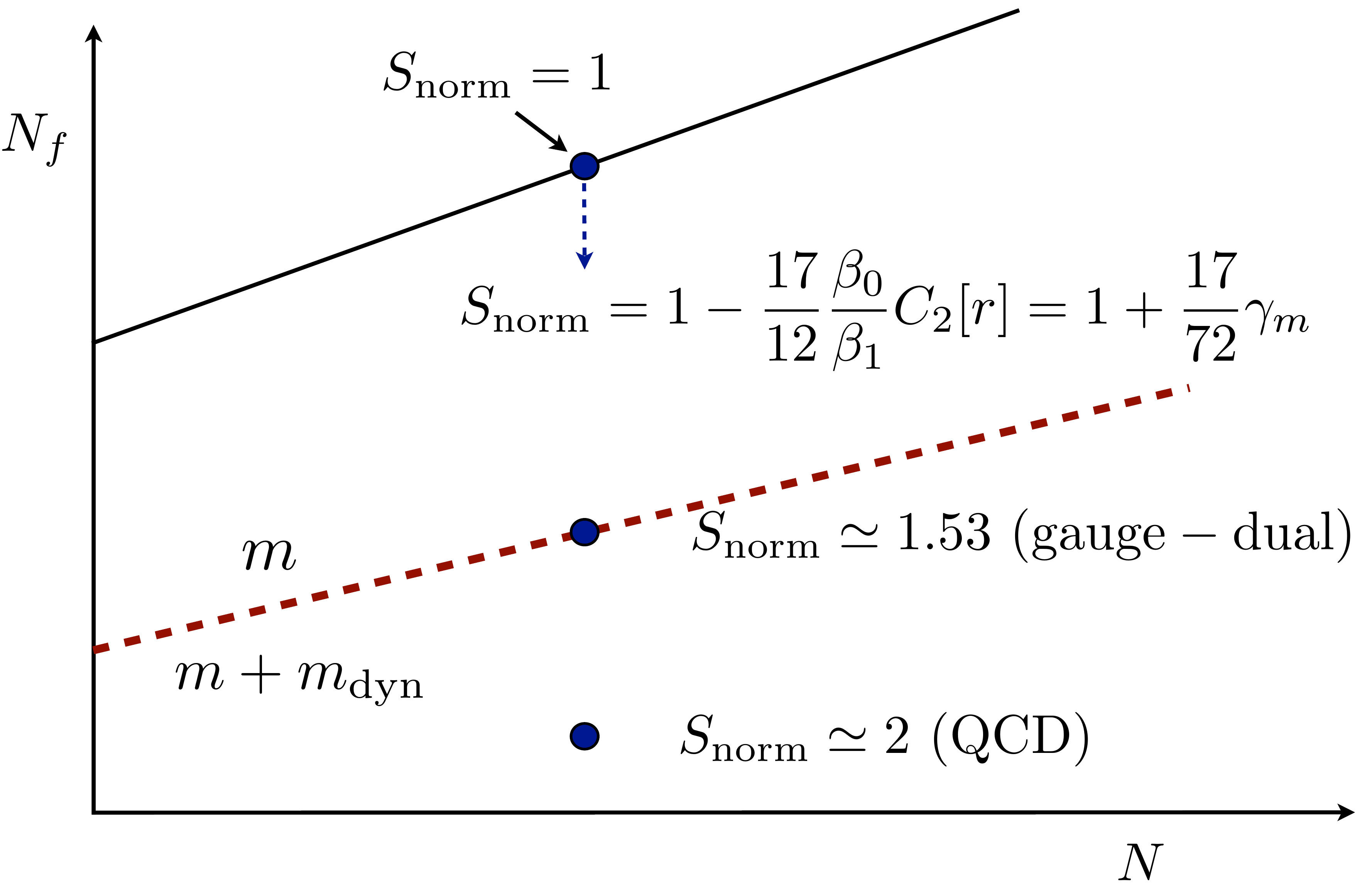} 
\caption{Cartoon of the dependence of the normalized $S$-parameter ($S_{\rm norm}$) on the number of Dirac flavors transforming according to the fundamental representation of the $SU(3)$ gauge theory across the phase diagram. The solid oblique line corresponds to the points where the theory looses asymptotic freedom. Chiral symmetry breaks below the dashed line while the conformal window is between the two lines. $S_{\rm norm} =1$ at the upper end of the conformal window and it increases according to the formulae    \eqref{Sgamma}  when decreasing the number of flavors. This result is exact within the perturbative regime. The estimate at the lower end of the conformal window has been derived using gauge duality in \cite{Sannino:2010fh}.    The QCD value is reported too. Below the conformal window a dynamical mass is generated (on the top of the bare mass $m$) and it is expected to vanish smoothly across the lower boundary suggesting that the $S$-parameter is smooth too.} 
\label{S-cartoon}
\end{center}
\end{figure}

Therefore we expect the lower bound on the normalized $S$-parameter to apply to the entire phase diagram concerning asymptotically free gauge theories.  We elucidate the above picture in Fig.~\ref{S-cartoon}.

The presence of a lower bound does not contradict  the statements made earlier in the literature that the $S$-parameter in near conformal theories can be smaller than the one in QCD \cite{Appelquist:1998xf, Kurachi:2006mu,Kurachi:2006ej,Kurachi:2007at}. A reduction of $S_{\rm norm}$ with respect to the QCD value is possible but should not violate the bound \eqref{sbound} suggested in \cite{Sannino:2010ca}. In particular we do not expect a negative $S$-parameter to occur in an asymptotically free gauge theory. While we work in a controlled regime in which our prediction for the flavor dependence of the $S$-parameter is exact we note that such a dependence has been long sought after. In fact many estimates have been provided in the literature using various approximations in field theory  \cite{Sundrum:1991rf}  or using computations inspired by the original AdS/CFT correspondence \cite{Maldacena:1997re} in \cite{Hong:2006si,Hirn:2006nt,Piai:2006vz,Agashe:2007mc,Carone:2007md,Hirayama:2007hz}. Recent attempts to use AdS/CFT inspired methods can be found in \cite{Dietrich:2009af,Dietrich:2008up,Dietrich:2008ni,Hirn:2008tc,Nunez:2008wi,Fabbrichesi:2008ga,Anguelova:2010qh}.

The present results have a strong impact on the construction of models of dynamical electroweak symmetry breaking. In fact they show that one family technicolor models are strongly disfavored with respect to precision data. However walking technicolor models with the smallest number of techniflavors  gauged under the electroweak symmetry are favored by precision tests \cite{Sannino:2004qp, Hong:2004td,Dietrich:2005jn,Dietrich:2005wk,Evans:2005pu,Christensen:2005cb,Gudnason:2006ug,Gudnason:2006yj,Foadi:2007ue,Foadi:2008xj,Belyaev:2008yj,Foadi:2008ci,Foadi:2007se}. These include models of partially gauged technicolor \cite{Dietrich:2005jn,Dietrich:2006cm,Fukano:2009zm,Frandsen:2009fs} in which only two techniflavors are electroweak gauged. 

The generalization to symplectic and orthogonal technicolor gauge groups \cite{Sannino:2009aw} is straightforward and the results interesting since {\it orthogonal} technicolor models \cite{Frandsen:2009mi} have already been proposed in the literature. 

The results obtained in the limit of sending to zero the mass of the fermions at a nonzero external momentum is also interesting since it applies immediately to models of unparticle physics with unparticle matter gauged under the weak interactions.  

\vskip .5cm
\section{Minimal Technicolor}
The simplest models of TC which are shown to pass the precision tests were put forward recently \cite{Sannino:2004qp,Dietrich:2005jn}.  To be concrete we describe here the MWT \cite{Sannino:2008ha} extension of the SM.

 The gauge group is $SU(2)_{TC}\times SU(3)_C\times SU(2)_L\times U(1)_Y$ and the field content of the technicolor sector is constituted by four techni-fermions and one techni-gluon all in the adjoint representation of $SU(2)_{TC}$.  The model features also a pair of Dirac leptons, whose left-handed components are assembled in a weak doublet, necessary to cancel the Witten anomaly \cite{Witten:1982fp} arising when gauging the new technifermions with respect to the weak interactions. Summarizing, the fermionic particle content of the MWT is given explicitly by
\beq Q_L^a=\left(\begin{array}{c} U^{a} \\D^{a} \end{array}\right)_L , \qquad U_R^a \
, \quad D_R^a, \quad a=1,2,3 \ ; \quad
L_L=\left(
\begin{array}{c} N \\ E \end{array} \right)_L , \qquad N_R \ ,~E_R \ . 
\label{MWTp}
\eeq 
The following generic hypercharge assignment is free from gauge anomalies:
\begin{align}
Y(Q_L)=&\frac{y}{2} \ ,&\qquad Y(U_R,D_R)&=\left(\frac{y+1}{2},\frac{y-1}{2}\right) \ ,  \nonumber\\
Y(L_L)=& -3\frac{y}{2} \ ,&\qquad
Y(N_R,E_R)&=\left(\frac{-3y+1}{2},\frac{-3y-1}{2}\right) \ \label{assign2} .
\end{align}
The global symmetry of this TC theory is $SU(4)$, which breaks explicitly to $SU(2)_L \times U(1)_Y$ by the natural choice of the EW embedding \cite{Sannino:2004qp,Dietrich:2005jn}. EWSB is triggered by a fermion bilinear condensate and the vacuum choice is stable against the SM quantum corrections \cite{Dietrich:2009ix}. This model can feature very light composite Higgs scalars as advocated in \cite{Hong:2004td,Dietrich:2005jn,Dietrich:2006cm} and  supported by the recent analysis performed by Natale's group \cite{Doff:2009na,Doff:2009kq,Doff:2009nk,Doff:2007zz,Doff:2008xx,Doff:2005vu}. 

An explicit construction of an extended TC type model  addressing the problem of giving mass to the third generation of quarks and the new generation of leptons appeared in \cite{Evans:2005pu}. 

A less natural model introducing a novel scalar ({\it bosonic TC}) mimicking the effects of the extended TC interactions has also been introduced in \cite{Antola:2009wq,Zerwekh:2009yu} following the pioneering work of Simmons  \cite{Simmons:1988fu},  Kagan and Samuel \cite{Kagan:1991gh}, and Carone  \cite{Carone:1992rh,Carone:1994mx}. More recently this type of models have been investigated also in \cite{Hemmige:2001vq,Carone:2006wj,Zerwekh:2009yu}. Interesting related work can be also found in \cite{Chivukula:1990bc,Chivukula:2009ck}.  Eventhough these models are phenomenologically viable, they suffer from a SM-like fine tuning and are therefore unnatural. Supersymmetric TC has been considered \cite{Dine:1981za,Dobrescu:1995gz} as a way to naturalize bosonic TC. Another possibility would be to imagine the new scalars also to be composite of some new strong dynamics.

\section{Minimal Supersymmetric Conformal Technicolor: Unification in Theory Space}

We investigate now the supersymmetrized version of the MWT \cite{Antola:2010nt,Antola:2010jk}. The resulting model provides a natural ultraviolet completion of the model introduced in \cite{Antola:2009wq}. The appeal of supersymmetry (SUSY) resides in its higher level of space-time symmetries as well as in its often praised natural link to string theory. The most investigated route to introduce SUSY has been to supersymmetrize the SM and then invoke some mechanism to break SUSY again, given that no sign of the superpartners has yet been observed in experiments.

We fuse the basic features of TC and SUSY to construct a natural supersymmetric TC theory possessing several interesting theoretical and phenomenological features. Besides symmetry arguments another equally relevant reason to look for a supersymmetric TC extension of the SM  is linked to the fact that the generation of the SM fermion masses is less involved than in models with total absence of scalars, although still natural.

The supersymmetric TC idea was put forward in \cite{Dine:1981za}, though the phenomenological viability of these early models seemed difficult to achieve. An important difference with our model is that the underlying supersymmetric and TC theories, which can be resumed by decoupling, either the TC fields or the superpartners by sending their masses to infinity, are both phenomenologically viable.

The basic properties of the model are: 
\begin{itemize}
\item The model possesses the highest degree of four-dimensional space-time symmetry compatible with experiments.
\item The model can interpolate between already studied extensions of the SM at the TeV scale, such as unparticle physics \cite{Georgi:2007ek,Georgi:2007si},  MWT \cite{Sannino:2004qp,Dietrich:2005jn}, and MSSM (see \cite{Martin:1997ns} for a review). Hence the models can be used as a well defined laboratory to investigate different theoretical ideas. 
\item The model is a UV complete theory in which fermions naturally acquire mass. 
\item The models possess a clear and direct link to string theory in such a way that AdS/CFT techniques \cite{Maldacena:1997re} are readily applicable to realistic extensions of the SM. 
\end{itemize}

\begin{figure}
\begin{center}
\includegraphics[width=16cm]{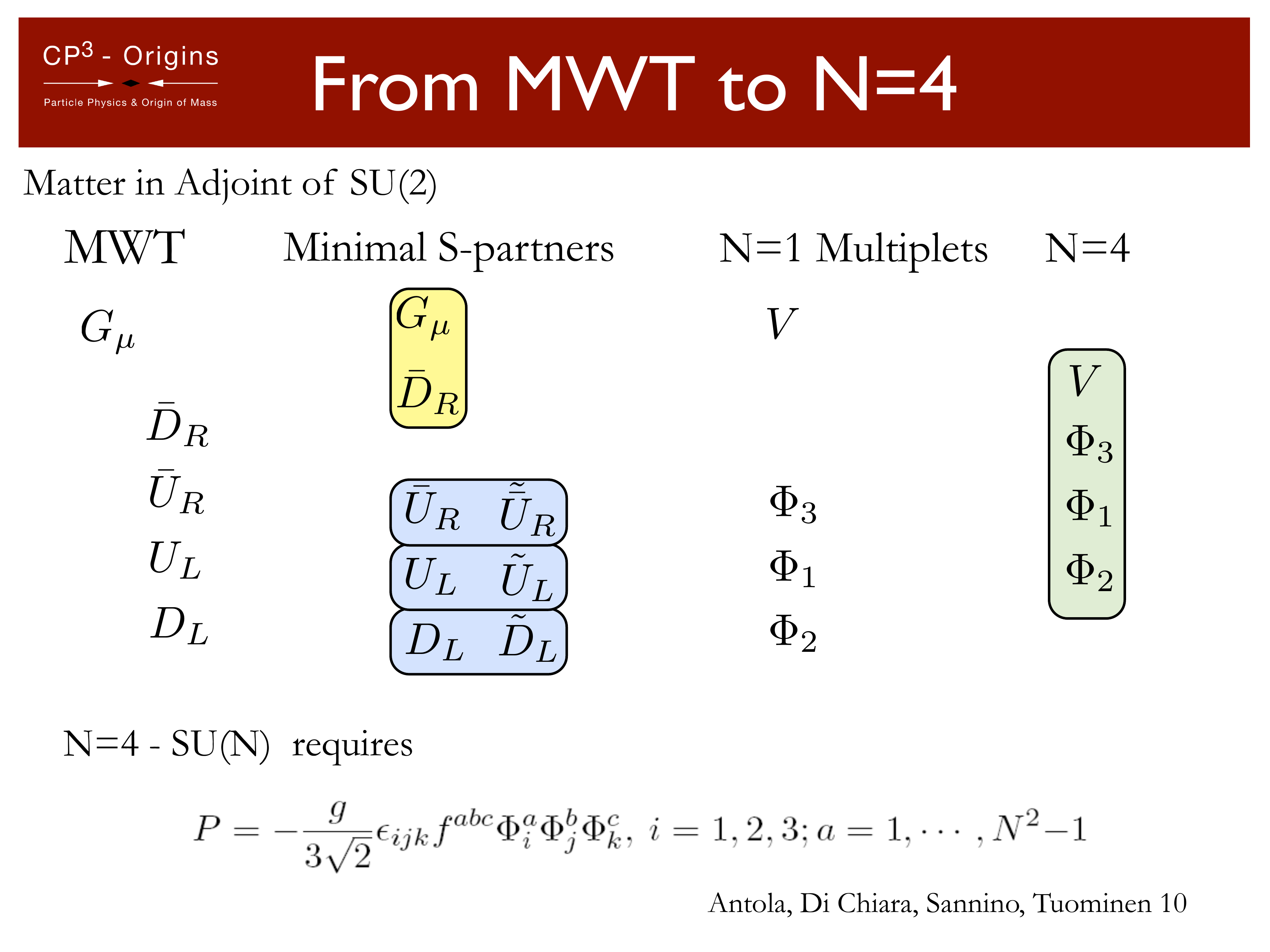}
\end{center}
\label{MSCT}
\caption{We start with the MWT gauge theory, first column on the left, and unite the first Weyl fermion transforming according to the adjoint representation of the TC gauge group with the technigluons forming the Vector superfield $V$, we then add 3 complex scalars and together with the remaining 3 Weyl fermions form 3 Chiral superfields, all transforming according to the adjoint representation of the gauge group. In the last column we make manifest the fact that the Vector and three Chiral superfields amount to the ${\cal N}=4$ supermuliplet. At the bottom of the slide the required  ${\cal N}=4$ superpotential is shown. }
\end{figure}

We start with the observation that the fermionic and gluonic spectrum of MWT fits perfectly in an ${\cal N}=4$ supermultiplet,  provided that we also include three scalar superpartners. In fact the $SU(4)$ global symmetry of MWT is nothing but the well known $SU(4)_R$ $R$ symmetry of the ${\cal N}=4$ Super Yang-Mills (4SYM) theory. Having at hand already a great deal of the spectrum of 4SYM we explore the possibility of using this theory as a natural candidate for supersymmetric TC. We show in Fig.~\ref{MSCT}, from left to right, step by step how to naturally embed MWT into the famous 4SYM theory.

We gauge part of the $SU(4)_R$ global symmetry of the supersymmetric TC theory in order to couple the new supersymmetric sector to the weak and hypercharge interactions of the SM. We choose to do this in such a way that the model can still preserve ${\cal N} =1 $ SUSY. To this end one of the four Weyl technifermions is identified with the techni-gaugino and should be a singlet under the SM gauge group. The only possible candidates for this role are $\bar{U}_R$ and $\bar{D}_R$, for $y=\mp 1$ respectively: we arbitrarily choose $y=1$ and identify $\bar{D}_R$ with the techni-gaugino. With this choice the charge assignments of the particles in Eq.(\ref{MWTp}) under $SU(2)_{TC}\times SU(3)_C\times SU(2)_L\times U(1)_Y$ are
\be
&&Q_L\sim \left(\textbf{3},\textbf{1},\textbf{2},\frac{1}{2}\right),\ \bar{U}_R\sim(\textbf{3},\textbf{1},\textbf{1},-1),\ \bar{D}_R\sim(\textbf{3},\textbf{1},\textbf{1},0),\ \nonumber\\
&&L_L\sim\left(\textbf{1},\textbf{1},\textbf{2},-\frac{3}{2}\right),\ \bar{N}_R\sim(\textbf{1},\textbf{1},\textbf{1},1),\ \bar{E}_R\sim(\textbf{1},\textbf{1},\textbf{1},2). 
\label{chaa}
\ee
Based on these assignments we then define the scalar and fermion components of the ${\cal N}=4$ superfields via 
\beq
\left(\tilde{U}_L,\ U_L\right)\in \Phi_1,\quad \left(\tilde{D}_L,\ D_L\right)\in \Phi_2,\quad   
\left(\tilde{\bar{U}}_R,\ \bar{U}_R\right)\in \Phi_3,\quad \left(G,\ \bar{D}_R\right)\in V,  
\label{superq}
\eeq
where we used a tilde to label the scalar superpartner of each fermion. We indicated with $\Phi_i$, $i=1,2,3$ the three chiral superfields of 4SYM and with $V$ the vector superfield.  Four more chiral superfields are necessary to fully supersymmetrize the MWT model, i.e.:
\beq
\left(\tilde{N}_L,\ N_L\right)\in \Lambda_1,\quad \left(\tilde{E}_L,\ E_L\right)\in \Lambda_2,\quad   
\left(\tilde{\bar{N}}_R,\ \bar{N}_R\right)\in N,\quad\left(\tilde{\bar{E}}_R,\ \bar{E}_R\right)\in E.
\label{superl}
\eeq

As one can see from the spectrum in Eq.(\ref{chaa})  there is no scalar field that can be coupled to SM fermions in a gauge invariant way and play the role of the SM Higgs boson (a weak doublet with hypercharge $Y=\pm\frac{1}{2}$). We therefore 
(re)introduce in the theory two Higgs doublet superfields with respective charge assignment
\beq
H\sim \left(\textbf{1},\textbf{1},\textbf{2},\frac{1}{2}\right),\ H^{\prime}\sim\left(\textbf{1},\textbf{1},\textbf{2},-\frac{1}{2}\right),
\label{superH}
\eeq
where the presence of both $Y=\pm\frac{1}{2}$ superfields is needed to give mass by gauge invariant Yukawa terms to both the upper and lower components of the weak doublets of SM fermions. With this choice it is rather natural to take the MSSM to describe the supersymmetric extension of the SM sector. All the MSSM fields are defined as singlets under $SU(2)_{TC}$. The resulting MSCT model is naturally anomaly-free, since both the MWT and the MSSM are such. 
We summarize in Table \ref{MSCTsuperfields} the quantum numbers of the superfields in Eqs.(\ref{superq},\ref{superl},\ref{superH}).
\begin{table}[h]
\centering
\begin{tabular}{c|c|c|c|c}
\noalign{\vskip\doublerulesep}
Superfield & SU$(2)_{TC}$ & SU$(3)_{\text{c}}$ & SU$(2)_{\text{L}}$ & U$(1)_{\text{Y}}$\tabularnewline[\doublerulesep]
\hline
\noalign{\vskip\doublerulesep}
$\Phi_{1,2}$ & Adj & $1$ & $\square$ & 1/2\tabularnewline[\doublerulesep]
\noalign{\vskip\doublerulesep}
$\Phi_{3}$ & Adj & 1 & 1 & -1\tabularnewline[\doublerulesep]
\noalign{\vskip\doublerulesep}
$V$ & Adj & 1 & 1 & 0\tabularnewline[\doublerulesep]
\noalign{\vskip\doublerulesep}
$\Lambda_{1,2}$ & 1 & 1 & $\square$ & -3/2\tabularnewline[\doublerulesep]
\noalign{\vskip\doublerulesep}
$N$ & 1 & 1 & 1 & 1\tabularnewline[\doublerulesep]
\noalign{\vskip\doublerulesep}
$E$ & 1 & 1 & 1 & 2\tabularnewline[\doublerulesep]
\noalign{\vskip\doublerulesep}
$H$ & 1 & 1 & $\square$ & 1/2\tabularnewline[\doublerulesep]
\noalign{\vskip\doublerulesep}
$H'$ & 1 & 1 & $\square$ & -1/2\tabularnewline[\doublerulesep]
\end{tabular}\caption{MSCT ${\cal N}=1$ superfields}
\label{MSCTsuperfields}
\end{table}
The renormalizable superpotential for the MSCT, allowed by gauge invariance, and which we require additionally to be ${\cal N}=4$ invariant in the limit of $g_{TC}$ much greater than the other coupling constants and to conserve baryon and lepton numbers is
\beq
P=P_{MSSM}+P_{TC},
\label{spot}
\eeq
where $P_{MSSM}$ is the MSSM superpotential, and
\beq
P_{TC}=-\frac{g_{TC}}{3\sqrt{2}} \epsilon_{ijk} \epsilon^{abc} \Phi^a_i \Phi^b_j \Phi^c_k+y_U \epsilon_{ij3}\Phi^a_i H_j\Phi^a_3+y_N \epsilon_{ij3}\Lambda_i H_j N+y_E \epsilon_{ij3}\Lambda_i H^{\prime}_j E+y_R \Phi^a_3 \Phi^a_3 E.
\label{spmwt}
\eeq
This superpotential describes an approximately conformal theory in the limit when  $g_{TC}$ is much greater than the gauge couplings $g_Y,\ g_L$, and Yukawa couplings $y_U,\, y_N,\, y_E,\, y_R$. It is interesting to notice that a generic Yukawa coupling constant $y_{TC}$ in the first term has a infrared fixed point at $g_{TC}$ \cite{Petrini:1997kk}, therefore justifying our choice of taking $y_{TC}=g_{TC}$.

The Lagrangian of the MSCT is
\beq
{\cal L}={\cal L}_{MSSM}+{\cal L}_{TC} \ ,
\label{smwt}
\eeq
where the supersymmetric TC Lagrangian can be written as:
\beq
{\cal L}_{TC}=\frac{1}{2} \tr \left(W^{\alpha} W_{\alpha}|_{\theta\theta}+\bar{W}_{\dot{\alpha}} \bar{W}^{\dot{\alpha}}|_{\bar{\theta}\bar{\theta}}\right)+\Phi_f^{\dagger}\exp \left( 2 g_X V_X \right) \Phi_f|_{\theta\theta\bar{\theta}\bar{\theta}}+\left(P_{TC}|_{\theta\theta}+h.c.\right),
\label{tecL}
\eeq
where
\beq
\Phi_f=Q,\Phi_3,\Lambda,N,E;\quad X=TC, C, L, Y\ ,
\eeq
with $Q$ and $\Lambda$ defined as the weak doublet superfields with components $\Phi_1,\ \Phi_2$, and $\Lambda_1,\ \Lambda_2$, respectively. The product $g_X V_X$ is assumed to include the gauge charge of the superfield on which it acts. The charge is $Y$ for $U(1)_Y$, and 1 (0) for a multiplet (singlet) of a generic group $SU(N)$. The TC vector superfield $V_{TC}$ is identified with $V$ and its elementary field components are given in Eq.(\ref{superq}). The remaining vector superfields are those already defined in the MSSM \cite{Martin:1997ns} while the superpotential $P_{TC}$ is given in Eq.(\ref{spmwt}). Finally, the first term on the right of Eq.(\ref{tecL}) and its Hermitian conjugate represent the kinetic Lagrangian terms of the self-interacting techni-gluon and techni-gaugino.

\begin{figure}
\begin{center}
\includegraphics[width=16truecm]{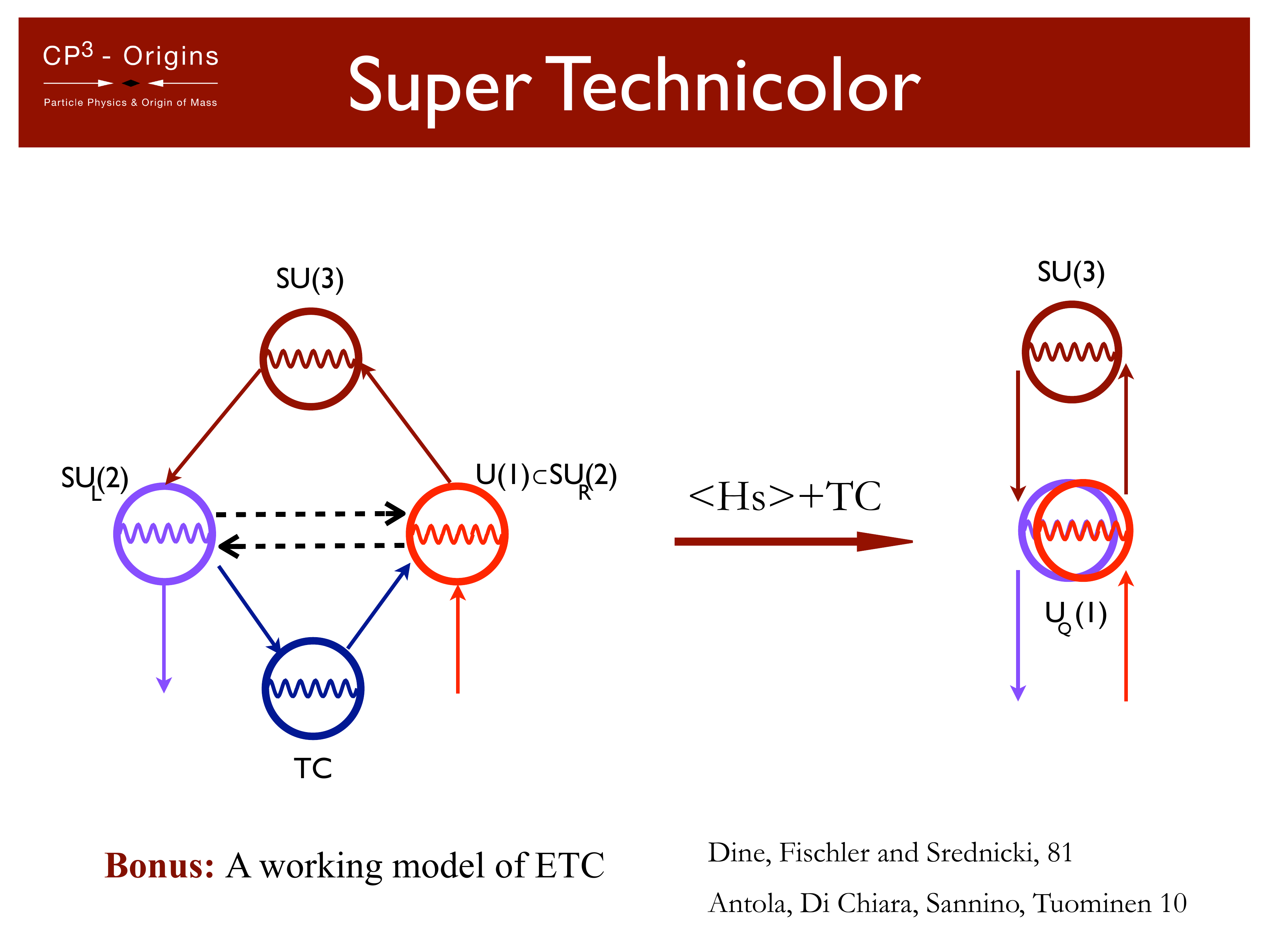}
\end{center}
\label{SuperTC}
\caption{We show a simplified cartoon of the unification of several models in theory space.  Each wavy-line corresponds to Vector superfields and each straight line a Chiral supefield.  In the picture one observes the simultaneous presence of the MSSM Higgses (the two dashed lines) and of the superymmetric TC sector. }
\end{figure}

The unification in theory space is made more clear in Fig.~\ref{SuperTC} in where besides the fact that each line and blob is correctly supersymmetrized one observes the simultaneous presence of the MSSM Higgses (the two dashed lines) and of the super technicolor sector. 

\subsection{The MSCT Landscape}


One of the features of the MSCT spectrum is the existence of a supersymmetric fourth family of leptons. The MWT predicts the natural occurrence of a fourth family of leptons around the EW energy scale, put forward first in \cite{Dietrich:2005jn}. The physics of these fourth family of leptons has been studied in \cite{Antipin:2009ks,Frandsen:2009fs}. From the EW  point of view there is little difference between the MWT and a fourth-family extended SM at the EW scale. 

Since the MSCT is a supersymmetrized version of the MWT the former now features, besides the techniquarks, a novel and natural super 4th family of leptons awaiting to be discovered at colliders, albeit with more exotic electric charges: the new electron will be doubly charged and will have a number of interesting signatures at colliders. 

Several further features of the spectrum are dependent on which part of the MSCT parameter space one chooses to study. The MSCT allows model builders to investigate a large number of (perturbative and nonperturbative) inequivalent extensions of the (MS)SM. These inequivalent extensions are determined, partially, by the choice of the value of the coupling constant $g_{TC}$ of the supersymmetric TC sector near the EW scale as well as by the vacuum choice permitted by the flat directions and, finally, by the SUSY breaking pattern. We now discuss two different limits one can take within the MSCT and sketch some of their basic features; each specific model deserves to be studied on its own and some of these models will be investigated in  more detail in future publications. 

The simplest case to consider is the one in which the new sector is weakly coupled at the EW scale and can be treated perturbatively. In this case the spectrum of states, which can be observed at the EW scale, is constituted by the elementary fields introduced in \eqref{superq} and \eqref{superl}, plus the MSSM ones. However, the detailed mass spectrum will depend on the structure of the SUSY breaking terms and on the corrections induced by the EW symmetry on the supersymmetric TC sector. 

The spectrum is extremely rich with several novel weakly coupled particles, such as the new techni-up and techni-down, and their respective superpartners, which can emerge at the LHC. The superpartners will be very similar to ordinary squarks but will carry TC instead of color. All the weak processes involving the production of squarks at colliders should be re-investigated to take into account the presence of these new states.

 \begin{figure}[t]
\begin{center}
\includegraphics[width=16truecm]{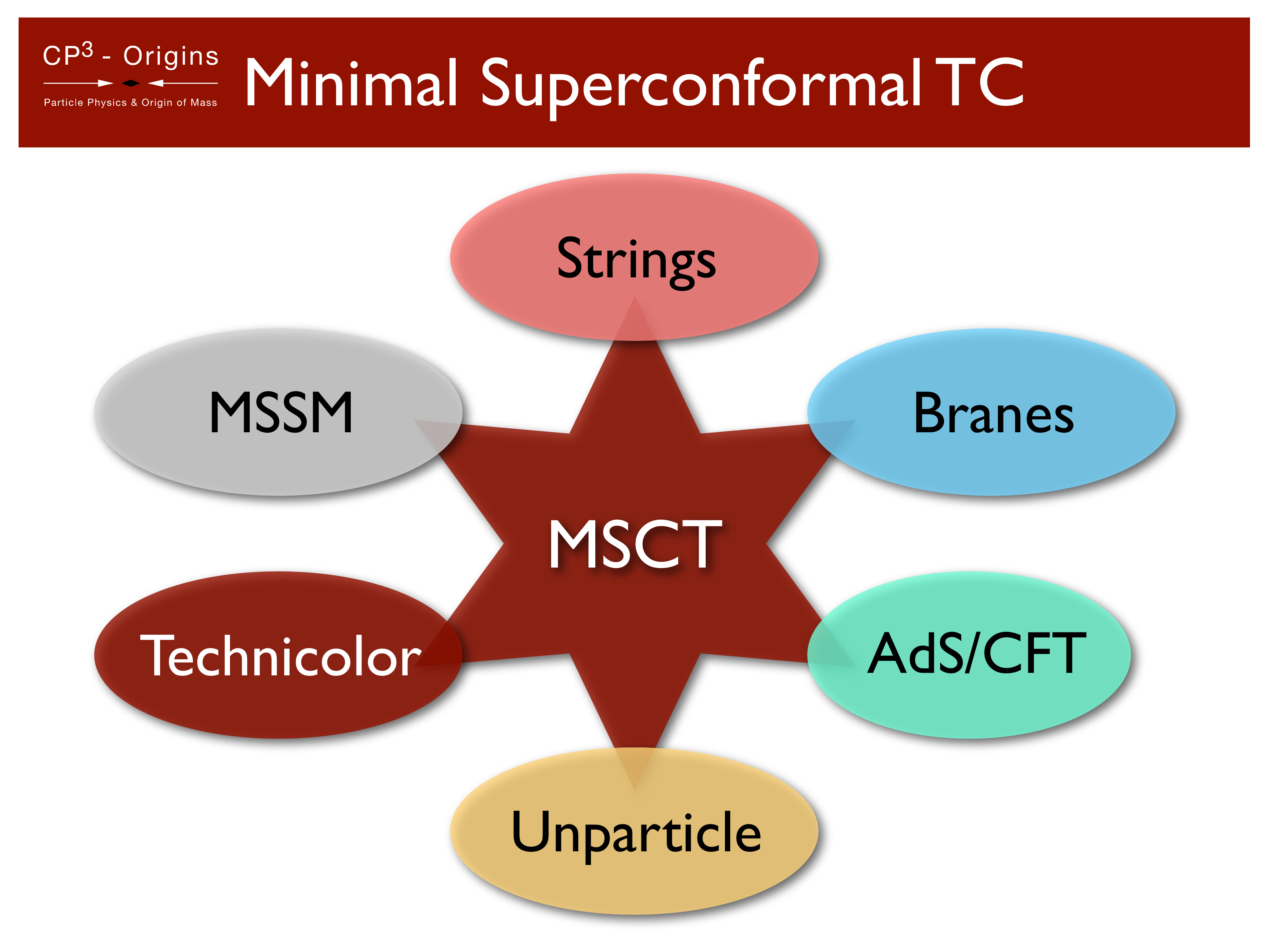}
\end{center}
\caption{Schematic representation of the fact that MSCT unifies different natural extensions of the SM models, depending on the parameters chosen within the model.}
\label{unificationTS}
\end{figure} 

If  we assume the supersymmetric TC dynamics to be strongly coupled at the EW scale, then we must use non-perturbative methods to investigate the effects of the new sector on the MSSM dynamics and vice versa. {}For example, we can no longer  use the single particle state interpretation in terms of the underlying degrees of freedom of the supersymmetric technicolor model but rather must use an {\it unparticle} language given that the supersymmetric TC model is exactly conformal, before coupling it to the MSSM.  The model resembles the one proposed in \cite{Sannino:2008nv} in which, besides a TC sector, one has also coupled a natural unparticle composite sector. If no SUSY breaking terms are added directly to the 4SYM sector then conformality will be broken only via weak and hypercharge interactions. A relevant point is that one can use the machinery of the AdS/CFT correspondence to make reliable computations in the nonperturbative sector, considering the effects of the EW interactions as small perturbations. 

 We have introduced a very minimal supersymmetric extension of MWT and shown that one can use ${\cal N}=4$  SYM as a direct extension of the SM of particle interactions. The model unifies different extensions of the SM, from unparticle physics to supersymmetry and technicolor as schematically represented in Fig.~\ref{unificationTS}

 \subsection*{Acknowledgments}
Part of the material reviewed here is based on work done in collaboration with Matti Antola,  Stefano di Chiara, Dennis D. Dietrich, Roshan Foadi, Mads T. Frandsen, Hidenori Fukano-Sakuma, Matti J\"arvinen, Claudio Pica, Thomas A. Ryttov and Kimmo Tuominen.  

I would like to thank the organizers of the conference for having provided a vibrant scientific atmosphere.

\end{document}